**Electrical spin injection into graphene through monolayer hexagonal boron nitride**

Takehiro Yamaguchi[1], Yoshihisa Inoue[1], Satoru Masubuchi[1,2], Sei Morikawa[1], Masahiro Onuki[1], Kenji Watanabe[3], Takashi Taniguchi[3], Rai Moriya[1,*)], and Tomoki Machida[1,2,4, †)]

[1] *Institute of Industrial Science, University of Tokyo, 4-6-1 Komaba, Meguro, Tokyo 153-8505, Japan*

[2] *Institute for Nano Quantum Information Electronics, University of Tokyo, 4-6-1 Komaba, Meguro, Tokyo 153-8505, Japan*

[3] *National Institute for Materials Science, 1-1 Namiki, Tsukuba 305-0044, Japan*

[4] *PRESTO, Japan Science and Technology Agency, 4-1-8 Honcho, Kawaguchi 332-0012, Japan*

We demonstrate electrical spin injection from a ferromagnet to a bilayer graphene (BLG) through a monolayer (ML) of single-crystal hexagonal boron nitride (h-BN). A $Ni_{81}Fe_{19}$/ML h-BN/BLG/h-BN structure is fabricated using a micromechanical cleavage and dry transfer technique. The transport properties across the ML h-BN layer exhibit tunnel barrier characteristics. Spin injection into BLG has been detected through non local magnetoresistance measurements.

*) E-mail address: moriyar@iis.u-tokyo.ac.jp
†) E-mail address: tmachida@iis.u-tokyo.ac.jp

Graphene and other families of two dimensional (2D) atomic crystals have been recognized as a new type of material system[1-3] and received much attention for spintronics applications[4]. Among them, single-layer graphene (SLG) and bilayer graphene (BLG) have already demonstrated a long spin relaxation time and spin diffusion length[5,6]. Electrical spin injection and detection have been demonstrated by inserting a tunnel barrier between a ferromagnetic layer and graphene[7-10]. However, the injection of highly spin-polarized carriers into these materials has not yet been demonstrated. The spin polarization achieved by electrical spin injection from a ferromagnet to SLG, BLG, or multilayer graphene (MLG) is around several tens of percent, far smaller than the value achieved in state-of-the-art magnetic tunnel junction (MTJ) devices. Modern MTJ devices with crystalline MgO tunnel barriers have been demonstrated to possess a large tunnel spin polarization exceeding 90%[11,12]. For graphene, the fabrication of a crystalline tunnel barrier by a conventional evaporation methods is a non trivial because there are no materials that can be easily evaporated on graphene, while also maintaining a close lattice match. An alternative tunnel barrier material and corresponding fabrication technique on the graphene surface are highly desired.

Recently, the fabrication of 2D atomic crystal tunnel barriers such as hexagonal boron nitride (h-BN), $MoS_2$, and $WS_2$ on graphene have been demonstrated and extensively studied by other groups[13-17]. These 2D crystals have several advantages for tunnel barrier applications. 1) Such materials can be exfoliated with a monolayer (ML)-thick resolution. 2) A single-crystalline flake can be fabricated. 3) A wide range of band gaps are available from these materials. Spin injection through these tunnel barriers is a fundamentally important subject for future graphene spintronics applications. However, thus far there are no reports on spin injection from a ferromagnet to graphene through a 2D crystal.

Amongst the many possible 2D materials, we choose hexagonal boron nitride (h-BN) as a tunnel barrier material for the spin injection into graphene. h-BN is well known as a crystalline substrate for graphene[18]. Carrier transport can be greatly improved by encapsulating graphene with h-BN[19]. h-BN has a wide-band-gap (~6 eV) and its lattice closely matches that of graphene. Very high quality single crystal h-BN is readily available[20]. Recently, a SLG/h-BN/SLG tunnel junction has been demonstrated by Britnell et al.[13] and revealed a pinhole-free character of the h-BN tunnel barrier. Moreover, recent first-principles studies predict that spin-polarized tunneling can be achieved in a ferromagnet/h-BN(001) hetero junction[21,22]. In this letter, we demonstrate the fabrication of a ML h-BN tunnel barrier between graphene and a ferromagnet using a micromechanical cleavage and dry transfer method. Electrical spin injection and detection has been demonstrated with non local magnetoresistance measurements up to 300 K. This study demonstrates a spin injection through a new type of single-crystal tunnel barrier on graphene.

A schematic illustration of the device structure and a scanning electron microscopy (SEM) image are shown in Figs. 1(a) and 1(b), respectively. BLG channel is encapsulated with the h-BN to improve carrier mobility and spin transport. At the same time, this structure enable us electrical spin injection from ferromagnetic permalloy (Py, $Ni_{81}Fe_{19}$) to BLG through the ML h-BN layer. First, we fabricate a ~20-nm-thick h-BN layer on a 300 nm $SiO_2/n^+$-Si(100) substrate using micromechanical cleavage. The thick h-BN layer acts as a substrate for graphene. On top of this structure, bilayer graphene (BLG) and a ML of h-BN is fabricated using a micromechanical cleavage and dry transfer method[18,23]. The thicknesses of the graphene and h-BN layers as measured by an atomic force microscopy are 0.70 and 0.34 nm, respectively. These thicknesses correspond to two and one MLs of graphene and h-BN, respectively. Finally, a 30 nm Py layer and a non magnetic 45 nm Au/6 nm Ti electrode are fabricated by standard electron beam (EB) lithography and EB evaporation. The doped Si substrate is used as a back gate for controlling the carrier concentration of the BLG. The width of the BLG is 1 μm and the width of the three ferromagnetic electrodes, Py1, Py2, and Py3, are 270, 580, and 380 nm, respectively. The distance between ferromagnetic electrodes Py2 and Py3 is 600 nm. The mobility of the BLG is determined as 2700 and 2300 $cm^2V^{-1}s^{-1}$ at 30 and 300 K, respectively.

We measured the *I-V* characteristics of the NiFe/ML h-BN/BLG junction using three-terminal measurements, i.e. current is applied from Py2 to Py1 and voltage is measured between Py3 and Au/Ti. The *I-V* curve at 30 K under various back gate voltages $V_G$ is shown in Fig. 2(a). A non linear *I-V* relationship was observed. The shape of the *I-V* curve weakly depends on $V_G$ within $V_G$ = -50–+50 V. After the measurement of several junction resistances, we obtained a resistance area product *RA* of the ML h-BN of 0.8–1.2 kΩ μm$^2$ near the zero bias voltage. The change in differential resistance (*dV/dI*) with respect to $I_{DC}$ for ML of the h-BN barrier is shown in Fig. 2(b) and reveals a 50% decrease in resistance at an $I_{DC}$ of ±50 μA (±30 mV). With increasing temperature, the junction resistance monotonically decreases and the ratio *RA* (1.6 K)/*RA* (300 K) is ~2 for most of the ML h-BN junction. The *RA* seems to scale with the number of h-BN layers, as shown in Fig. 2(c). The *RA* increases about a factor of a hundred from one to two MLs of h-BN. This behavior is quite similar to that reported for a h-BN tunnel barrier with a graphene electrode[13]. These transport measurements suggest that we successfully fabricated a ML-thick h-BN tunnel barrier between a ferromagnetic layer and BLG.

The non local magnetoresistance (NLMR) was measured to detect spin injection and transport in BLG. The measurement circuit is shown in Fig. 1(a). The in-plane magnetic field dependence of the NLMR measured at 30 K and 300 K with $V_G = +50$ V and $I_{DC} = +50$ μA is shown in Fig. 3(a). Note that the Dirac point (DP) of the BLG is located at $V_G=+7$ V. A clear NLMR is observed at both 30 and 300 K. This indicates electrical spin injection from ferromagnet to BLG through a ML-thick h-BN barrier. Since the spin-polarized carriers are injected from both ferromagnetic electrodes Py1 and Py2, the NLMR signal displays multiple resistance steps. The relative alignment between Py1, Py2, and Py3 is indicated by arrows in the figure. We define the amplitude of the NLMR, $\Delta R_{NL}$, as the difference in resistance between positions A and B in Fig. 2(a). The temperature dependence of $\Delta R_{NL}$ is shown in Fig. 3(b) measured at three different $V_G$ value. $\Delta R_{NL}$ shows a peak at low temperature. A similar feature was observed in previous experiments on SLG and MLG devices, and it is considered to be a consequence of either electron-phonon scattering[24] or the ferromagnetic contact[25]. Next, the Hanle effect is measured with magnetic field perpendicular to the sample, and the results are shown in Fig. 3(c). A clear Hanle effect signal is observed, which is additional evidence for electrical spin injection into the BLG through the h-BN layer. To eliminate any contribution from the Py1 electrode, we determined the Hanle curve using the following procedure. First, we measured the Hanle effect at the magnetization configuration corresponding to positions A and B in Fig. 2(a). Next, we subtract these two curves and divide the result by 2. The obtained Hanle curve can be analyzed using the following equation[26]:

$$R_{\mathrm{NL}} \propto \int_0^\infty \frac{1}{\sqrt{4\pi D_s t}} e^{-\frac{L^2}{4 D_s t}} \cos\left(\frac{g\mu_B B t}{\hbar}\right) e^{-\frac{t}{\tau_s}} dt, \qquad (1)$$

where $D_s$ is the spin diffusion constant, $\tau_s$ is the spin relaxation time, $L$ is the distance between the Py electrodes, $g$ is the electron $g$-factor, and $\mu_B$ is the Bohr magneton. The results of the fit are also shown in Fig. 3(c). The spin relaxation time and spin diffusion constant fit parameters are 55 ps and 0.034 m$^2$s$^{-1}$ at 30 K and 56 ps and 0.023 m$^2$s$^{-1}$ at 300 K, respectively. $\tau_s$ exhibits a small temperature dependence similar to the other graphene-based spin valve devices[6,9,24]. The spin diffusion length $\lambda_s$ is determined as 1.35 and 1.14 μm at 30 and 300 K, respectively. Considering these values together with the contact resistance and channel length, our device is within the contact-induced spin relaxation region[27]. In this region, the spin relaxation is dominated by spin absorption at the ferromagnetic electrode and thus makes it difficult for us to perform an accurate evaluation of the spin relaxation in the channel. A more reliable determination of $D_s$ and $\tau_s$ can be accomplished by fabricating devices with a longer channel length. This problem will be addressed in future experiments. In addition, since the resistance of the tunnel barrier scales exponentially with the h-BN thickness[13], the contact-induced spin relaxation can be eliminated with the current channel length if we use two or three MLs of h-BN as a tunnel barrier.

The difference in the NLMR $\Delta R$ between A and B as shown in Fig. 2(a) is measured at 30 K under various $V_G$ and $I_{dc}$ values, and the results are shown in Fig. 4(a). For comparison, we plot the $V_G$ dependence of the conductance $\sigma$ in Fig. 4(b). First, $\Delta R$ and $\sigma$ clearly coincide. This phenomenon can be observed when the contact resistance is in the so-called transparent regime[8]. Although the h-BN layer acts as a tunnel barrier for charge, for spins we must compare the junction resistance and the spin resistance of graphene $R_G = \rho_G \lambda_S / W_G$, where $\rho_G$ is the resistance and $W_G$ is the width of BLG. From this expression, the junction resistance is comparable to $R_G$, thus spin absorption effects at the ferromagnetic electrode significantly influences the NLMR. $R_G$ increases toward the DP due to the increase of $\rho_G$. Therefore, there is more spin absorption at the ferromagnetic layer and a lower $\Delta R$.

Next, $I_{dc}$ exhibits a much smaller effect on $\Delta R_{NL}$. This observation also supports the claim that $\Delta R$ is dominated by spin absorption at the ferromagnetic layer, which does not depend on $I_{dc}$. $\Delta R$ increases at positive (negative) $I_{dc}$ at a $V_G$ of +50 V (-50 V). This increase is due to the effect of carrier drift under the injector ferromagnet[28,29]. According to this model, carrier drift could suppress the spin absorption in the ferromagnetic electrode. The sign of this effect should be reversed when the carrier type is changed from electron to hole, and this can be seen in Fig. 4(a). We performed measurements from 1.6 to 300 K to change the ratio between the contact resistance and $R_G$. A similar $\Delta R$ vs. $V_G$ dependence was observed for all measurement temperatures.

In conclusion, we fabricated a ML crystalline h-BN tunnel barrier using a micromechanical cleavage and dry transfer technique. By this technique, we demonstrated spin injection into BLG through a ML h-BN. The spin injection efficiency is limited by the spin current absorption of the ferromagnet because of the low junction resistance of the ML h-BN barrier. Junction resistance can be increased by using a thicker h-BN layer as a tunnel barrier. Nevertheless, our study revealed that a one atom thick crystalline tunnel barrier could be used for electrical spin injection into graphene. The fabrication method presented here is unique compared to methods found in previous studies on lateral spin valve and MTJ devices. This new method should open up new possibilities for utilizing 2D atomic crystal barriers for spintronics.

This work was partly supported by PRESTO, the Japan Science and Technology Agency, Grants-in-Aid from the Ministry of Education, Culture, Sports, Science and Technology (MEXT), and the Project for Developing Innovation Systems of MEXT.

Figure captions

Fig. 1

(a) Schematic illustration of the device and non local measurement configuration. (b) Scanning electron microscopy (SEM) image of the fabricated device. False colors are used in this image for clarity.

Fig. 2

(a) Three-terminal *I-V* curve for ferromagnetic contact Py2 measured at 30 K with various $V_G$ values. The amplitude of $V_G$ is indicated by the color bar. (b) The current amplitude dependence of the junction resistance measured at $V_G = \pm 50$ V. (c) The ML number dependence of *RA* in a $Ni_{81}Fe_{19}$/ML h-BN/BLG junction.

Fig. 3

(a) Non local magnetoresistance (NLMR) loop measured at 300 and 30K with $V_G = +50$ V and $I_{SD} = +50$ µA. The solid arrows indicate the magnetization direction of Py1, Py2, and Py3. The dotted arrows indicate the sweep direction of the external magnetic field. (b) Temperature dependence of $\Delta R_{NL}$ measured at $I_{SD} = +50$ µA. (c) Hanle effect measured under the same conditions as in (a) with a perpendicular magnetic field.

Fig. 4

(a) $V_G$ dependence of the $\Delta R_{NL}$ measured at 30 K under various $I_{SD}$ values. (b) The $V_G$ dependence of the conductance $\sigma$ of the BLG channel.

Figure 1

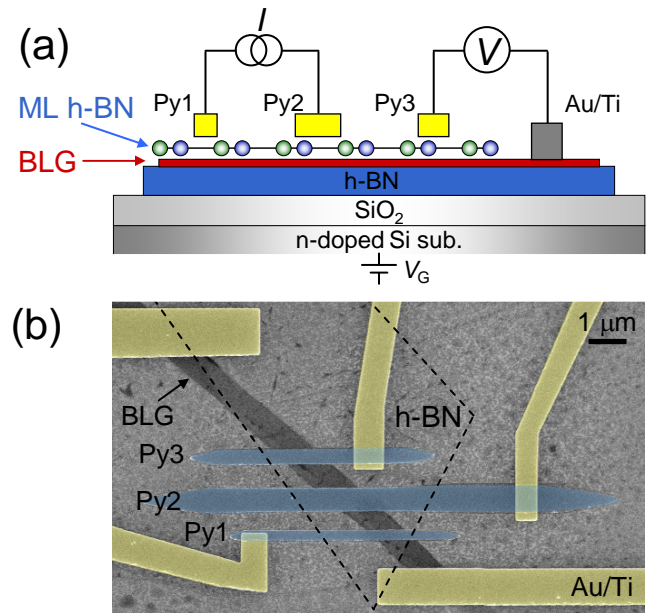

Figure 2

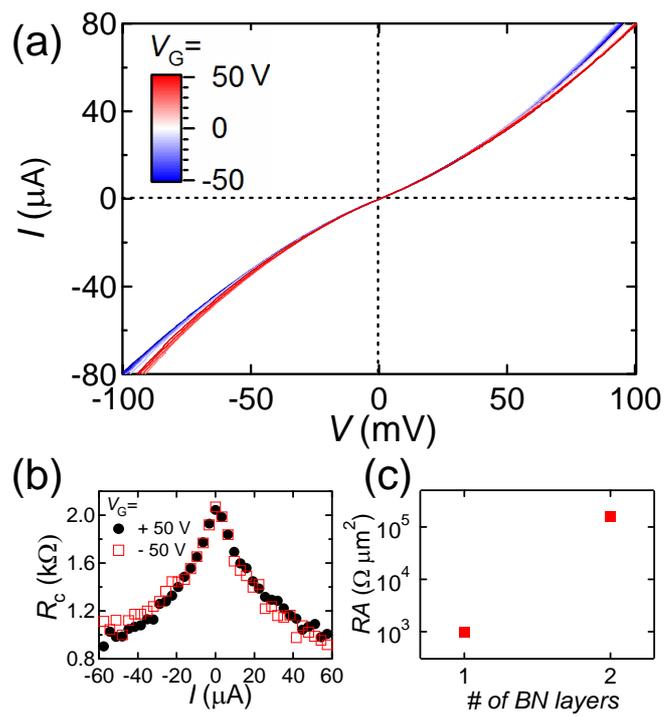

Figure 3

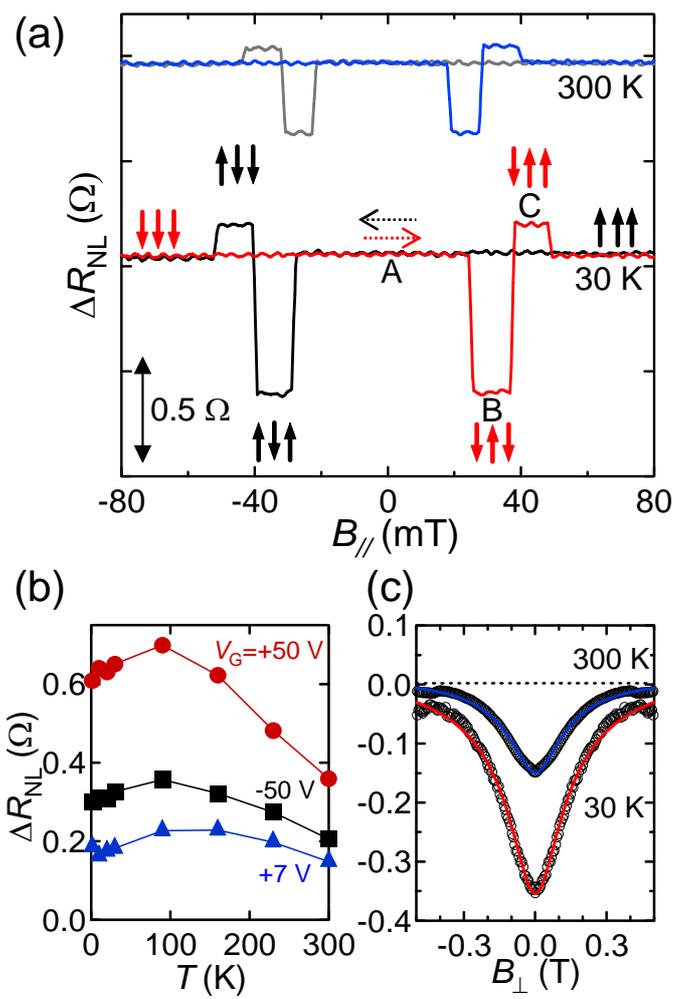

Figure 4

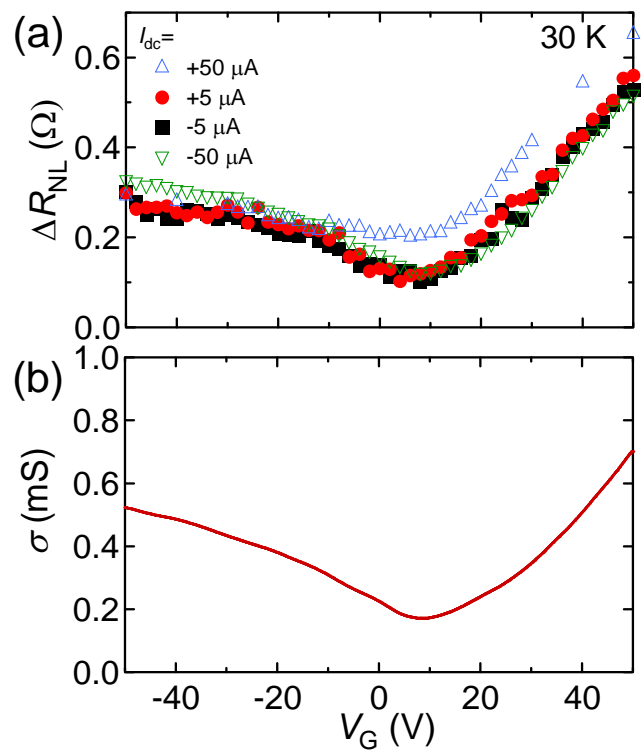